\documentclass[aps, prb, twocolumn, showpacs, groupedaddress, showkeys]{revtex4}
\usepackage[dvips]{color}
\usepackage{xcolor}
\usepackage{amsfonts}
\usepackage{amsmath}
\usepackage{amssymb}
\usepackage{graphicx}
\usepackage{mathrsfs}
\usepackage{bbm}
\usepackage{bm}
\usepackage{multirow}
\usepackage[sort&compress]{natbib}
\usepackage{dcolumn}%

\begin{document}

\newcommand{\dd}{d}
\newcommand{\pd}{\partial}
\newcommand{\myU}{\mathcal{U}}
\newcommand{\myr}{q}
\newcommand{\Urho}{U_{\rho}}
\newcommand{\myalpha}{\alpha_*}
\newcommand{\bd}[1]{\mathbf{#1}}
\newcommand{\Eq}[1]{Eq.~(\ref{#1})}
\newcommand{\Eqn}[1]{Eq.~(\ref{#1})}
\newcommand{\Eqns}[1]{Eqns.~(\ref{#1})}
\newcommand{\Figref}[1]{Fig.~\ref{#1}}
\newtheorem{theorem}{Theorem}
\newcommand{\me}{\textrm{m}_{\textrm{e}}}
\newcommand{\sgn}{\textrm{sign}}
\newcommand*{\bfrac}[2]{\genfrac{\lbrace}{\rbrace}{0pt}{}{#1}{#2}}

\newcommand{\CTensorName}{transport shape-averaged }

\title{Curved Space, curved Time, and curved Space-Time in Schwarzschild geodetic geometry}

\author{Rafael T. Eufrasio} 
\email{eufrasio@uark.edu }
\homepage{https://directory.uark.edu/people/eufrasio}

\affiliation{Department of Physics, University of Arkansas, Fayetteville, AR 72701} 

\author{Nicholas A. Mecholsky}
\email{nmech@vsl.cua.edu}

\affiliation{Department of Physics and Vitreous State Laboratory, 
The Catholic University of America,
Washington, DC 20064} 

\author{Lorenzo Resca}
\email{resca@cua.edu}
\homepage{https://physics.catholic.edu/}
\thanks{corresponding author.}

\affiliation{Department of Physics and Vitreous State Laboratory, 
The Catholic University of America,
Washington, DC 20064}

\date{\today}

\begin{abstract}

We investigate geodesic orbits and manifolds for metrics associated with Schwarzschild geometry, considering space and time curvatures separately. For `a-temporal' space, we solve a central geodesic orbit equation in terms of elliptic integrals. The intrinsic geometry of a two-sided equatorial plane corresponds to that of a full Flamm's paraboloid. Two kinds of geodesics emerge. Both kinds may or may not encircle the hole region any number of times, crossing themselves correspondingly. Regular geodesics reach a periastron greater than the $r_S$ Schwarzschild radius, thus remaining confined to a half of Flamm's paraboloid. Singular or $s$-geodesics tangentially reach the $r_S$ circle. These $s$-geodesics must then be regarded as funneling through the `belt' of the full Flamm's paraboloid. Infinitely many geodesics can possibly be drawn between any two points, but they must be of specific regular or singular types. A precise classification can be made in terms of impact parameters. Geodesic structure and completeness is conveyed by computer-generated figures depicting either Schwarzschild equatorial plane or Flamm's paraboloid. For the `curved-time' metric, devoid of any spatial curvature, geodesic orbits have the same apsides as in Schwarzschild space-time. We focus on null geodesics in particular. For the limit of light grazing the sun, asymptotic `spatial bending' and `time bending' become essentially equal, adding up to the total light deflection of 1.75 arc-seconds predicted by general relativity. However, for a much closer approach of the periastron to $r_S$, `time bending' largely exceeds `spatial bending' of light, while their sum remains substantially below that of Schwarzschild space-time.

\end{abstract}




\maketitle

\section{Introduction}  
The first and most fundamental solution of the field equations in Einstein's theory of general relativity (GR) was provided by Schwarzschild and published in 1916 in two ground-breaking papers.\cite{Schwarzschild(2016)} Schwarzschild's exact solution describes a static space-time in the vacuum outside a non-rotating spherical star or black-hole singularity at the origin. Geodesics in that space-time originally derived by Schwarzschild and Einstein have been widely studied and more deeply understood over time in various coordinate systems, as discussed in fundamental textbooks and review articles.\cite{Weinberg, MTW, Schutz2Ed, SchutzGravity, Wald, Rindler, Berry, Hobson, Hartle, Narlikar, Frolov, PricePrimer, MorrisThorne, Dadhich} There are two metrics closely associated with Schwarzschild's, which consider either space or time curvatures as separate from each other. Derivations and comparisons of geodesic orbit equations for all three metrics have been recently provided.\cite{Resca} In this paper we obtain exact solutions of all those geodesic orbit equations and analyze more deeply their manifolds. Remarkable results, both mathematically and physically, are presented. 

\section{Schwarzschild's Space-Time geometry and geodesics}\label{spacetime}
Schwarzschild's space-time geometry and metric line element

\begin{align}\label{metric}
ds^2  = & g_{\mu \nu} dx^{\mu} dx^{\nu} \nonumber\\
      = & -\bigg( 1- \frac{r_S}{r} \bigg)(c dt)^2 + \bigg(1 - \frac{r_S}{r} \bigg)^{-1} (dr)^2 + \nonumber\\
        & r^2 (d \theta)^2 + r^2 \sin^2 \theta (d \phi)^2 
\end{align}
are derived and discussed in fundamental GR textbooks, such as Refs.~\onlinecite{Weinberg, MTW, Schutz2Ed}. In \Eq{metric},
\begin{align}\label{radius}
r_{S} \equiv \frac{G}{c^2} 2 M
\end{align}
is Schwarzschild's radius. 

Four-momentum components $p^\mu = m \frac{d x^{\mu}}{d \tau}$ of material test particles have a \textit{time-like} pseudo-norm 
\begin{equation}\label{pseudo-norm}
  p_\mu p^\mu=g_{\mu \nu} p^\mu p^\nu =-m^2c^2 ,
\end{equation}
allowing $c{d \tau} = \sqrt{-ds^2}$ to represent an invariant proper-time interval.

A geodesic equation for four-momentum covariant components,
\begin{equation}\label{covariantgeodesic}
  m \frac{d p_{\beta}}{d \tau}
    = \frac{1}{2} \bigg( \frac{\partial g_{\nu \alpha}}{\partial x^{\beta}}  \bigg) p^\nu p^{\alpha},
\end{equation}
can be generally derived.\cite{Schutz2Ed} 

Conservation of energy, $m c^2 \tilde{E}$, and angular momentum, $m \tilde{L}$, lead to planar geodesic orbits, which can thus be assumed to be equatorial, with polar angle $\theta = \frac{\pi}{2} = \mathrm{const}$. We then arrive at a \textit{time-like geodesic orbit} equation in terms of the azimuthal angle, $\phi$, namely,
\begin{equation}\label{geodesicorbitradial}
  \bigg(\frac{d r}{d \phi}\bigg)^2
    = \frac{r^4}{\tilde{L}^2} \bigg\{c^2\tilde{E}^2 - c^2 +G\frac{2M}{r} - \frac{\tilde{L}^2}{r^2} + \frac{G}{c^2} \frac{2M}{r}   \frac{\tilde{L}^2}{r^2}  \bigg\}.
\end{equation}

In the non-relativistic limit we may omit the last term in \Eq{geodesicorbitradial} and rescale energy as to recover Newton's orbit equation.\cite{Resca}

We may also consider \textit{null geodesics}, having $ds^2=0$ in \Eq{metric}. These are traveled exclusively by massless test particles. Correspondingly, their four-momentum $p^\mu = \frac{d x^{\mu}}{d \lambda}$ has \textit{null} pseudo-norm
\begin{equation}
  p_\mu p^\mu= g_{\mu \nu} p^\mu p^\nu=0.
\end{equation}
Conservation of energy, $E$, and angular momentum, $L$, lead again to planar equatorial geodesics. The corresponding \textit{null geodesic orbit} equation is
\begin{equation}\label{nullgeodesicorbitradial}
  \bigg(\frac{dr}{d\phi}\bigg)^2 
    = \frac{r^4}{L^2}
      \bigg\{\frac{E^2}{c^2} -\frac{L^2}{r^2} + \frac{G}{c^2} \frac{2M}{r} \frac{L^2}{r^2} \bigg\}.
\end{equation}

\section{Proper spatial submanifold and geodesics in Schwarzschild submetric}\label{spatialschwarzschild}

Since Schwarzschild geometry is static, a natural way to separately consider proper space is to regard it as a three-dimensional (3D) submanifold at any given coordinate-time.\cite{O'Neill} This `fixed' or `a-temporal' space has a submetric line element for $x^{i} = (r, \theta, \phi)$ spatial coordinates
given by
\begin{align}\label{spatialmetric}
  d S^2 &= g_{ij} dx^i dx^j \nonumber\\
        &= \bigg(1 - \frac{r_S}{r} \bigg)^{-1} (dr)^2 + r^2 (d \theta)^2 + r^2 \sin^2 \theta (d \phi)^2.  
\end{align}

Up to the $r_S$ horizon, $d S^2 > 0$ represents the line element of a 3D positive-definite Riemannian submetric.
Therein, parameterizing geodesics with an affine parameter $\lambda$, tangent vectors $V^i=\frac{d x^i}{d \lambda}$ have a positive-definite norm 
\begin{equation}\label{spatialnorm}
  V_i V^i= g_{ij} V^i V^j = C^2 > 0 .
\end{equation}

Geodesic curves in the 3D spatial submanifold then obey the equation
\begin{equation}\label{spatialgeodesic}
  \frac{d V_k}{d \lambda}= \frac{1}{2} \bigg( \frac{\partial g_{ij}}{\partial x^k} \bigg) V^i V^j .
\end{equation}
Spherical symmetry leads again to planar geodesic curves, which can be thus assumed to be equatorial. Conservation of an  angular momentum equivalent, $L$, leads to a geodesic equation for the radial curvature coordinate, namely,
\begin{equation}\label{spatialradialgeodesic}
  \bigg(\frac{dr}{d\lambda}\bigg)^2
    = C^2- C^2 \frac{r_S}{r} -  \frac{L^2}{r^2} + \frac{r_S}{r}   \frac{L^2}{r^2}.
\end{equation}
From that, a geodesic \textit{orbit} equation, expressed in terms of the azimuthal angle, $\phi$, can be derived as
\begin{equation}\label{spatialradialgeodesicorbit}
\bigg( \frac{dr}{d \phi}  \bigg)^2= \frac{r^4}{L^2} \bigg\{ C^2- C^2 \frac{G}{c^2} \frac{2M}{r} -  \frac{L^2}{r^2} + \frac{G}{c^2}  \frac{2M}{r}   \frac{L^2}{r^2} \bigg\} .
\end{equation}

One may further consider weak-field and non-relativistic limits. In any case, it is clear that the exact time-like geodesic orbit \Eq{geodesicorbitradial} demands \textit{gravitational attraction} exclusively, whereas the exact spatial-submanifold geodesic orbit \Eq{spatialradialgeodesicorbit} invariably contains one term, namely its second, which corresponds to \textit{gravitational repulsion}.\cite{Resca}

It is particularly useful to generate a regular two-dimensional (2D) surface with curvature and metric equivalent to those of `a-temporal' space of Schwarzschild geodesic geometry. For the latter, we consider only a 2D subspace that represents geodesic equatorial planes. The corresponding line element in \Eq{spatialmetric} thus reduces to
\begin{equation}\label{reducedspatialmetric}
  d S^2= \bigg(1 - \frac{r_{S}}{r} \bigg)^{-1} (dr)^2 + r^2 (d \phi)^2  
\end{equation} 
for the $(r,\phi)$ coordinates. Then we embed the corresponding 2D submanifold in ordinary 3D Euclidean space, by associating $r^2$ with $(X^2 + Y^2)$ and by defining
\begin{equation}\label{Flamm}
  Z^2= 4 r_{S}^2\bigg(\frac{r}{r_{S}} - 1 \bigg).  
\end{equation} 
This is known as Flamm's paraboloid of revolution about the $Z-$axis.\cite{Flamm} It derives from straightforward integration after setting $d S^2= (d Z)^2 + (d r)^2 + r^2 (d \phi)^2$ equal to $d S^2$ in \Eq{reducedspatialmetric}. Thus Flamm's paraboloid is isometric to the 2D manifold of the geodesic equatorial plane within the Schwarzschild spatial submetric. Flamm's paraboloid originates most interesting dynamics of Einstein-Rosen bridge and wormhole constructions in Kruskal coordinates.\cite{MTW, Schutz2Ed, SchutzGravity, Wald, Rindler, Berry, Hobson, Hartle, Frolov, PricePrimer, MorrisThorne, EinsteinRosen} 

The geodesic orbit \Eq{spatialradialgeodesicorbit} admits a single turning point, obtained by equating \Eq{spatialradialgeodesicorbit} to its minimum zero value. One can then express the orbit periastron as  
\begin{equation}\label{periastron}
r_{p}^2 = {\frac{L^2}{C^2}} ,
\end{equation}
for any $r_{p} > r_{S}$. We may thus recast the orbit \Eq{spatialradialgeodesicorbit} solely in terms of $r_{p}$ and $r_{S}$ as 
\begin{equation}\label{spatialradialgeodesicorbitperiastron}
\bigg( \frac{dr}{d \phi}  \bigg)^2= \frac{r^4}{r_{p}^2} -  \frac{r_{S} r^3}{r_{p}^2} - r^2 + r_{S} r.
\end{equation} 

\section{Regular Geodesic orbits in curved space}\label{sec:geodesicorbits}
Let us also consider a four-dimensional (4D) pseudo-Riemannian manifold with metric 
\begin{align}\label{metricwithouttimecurvature} 
ds^2  = & g_{\mu \nu} dx^{\mu} dx^{\nu} \nonumber\\
      = & -(cdt)^2 + \bigg(1 - \frac{r_S}{r} \bigg)^{-1} (dr)^2 + \nonumber\\
        & r^2 (d \theta)^2 + r^2 \sin^2 \theta (d \phi)^2.
\end{align}
This differs from the Schwarzschild metric in that the time-like metric tensor component is assumed to be the same as it is in Special Relativity (SR), i.e., $g_{tt} = -1$, whereas the 3D spatial submanifold at any given coordinate time, $t$, maintains the same curvature, or $g_{rr}$, that it has in the Schwarzschild metric. 

Remarkably, all \textit{time-like, null} and \textit{space-like} geodesic orbit equations for this `splittable space-time' metric\cite{Price, Ellingson, Gruber} formally coincide with the \textit{space-like} geodesic orbit \Eq{spatialradialgeodesicorbit} that we derived for the `a-temporal' space of Schwarzschild geometry.\cite{Resca} It is possible to figure how that happens by keeping track of all $g_{tt}$ and $g_{rr}$ factors throughout the exact derivation of geodesic orbits for all metrics that we consider. A central element is that the product of $g_{tt}$ and $g_{rr}$ is constant only for the full space-time Schwarzschild metric. Maintaining $g_{tt}g_{rr}=-1$, as it is in Minkowski space-time, may indicate that time and space bend inversely, relative to each other, in Schwarzschild space-time. That may in turn reflect a basic requirement of the equivalence principle, namely, that the speed of light must remain a universal constant in any local freely-falling Lorentzian frame, in curved space-time of GR, as it is in flat space-time of SR. This result is peculiar to Schwarzschild coordinates, however. It extends only to a first order in $r_{S}/r$ in `isotropic coordinates,' as shown in Eq. (10.89) and p. 292 of Ref. \onlinecite{Schutz2Ed}, for example. In Flamm's coordinates, obtained from \Eq{Flamm} and \Eq{reducedspatialmetricFlamm}, the product of $g_{tt}$ and $g_{ZZ}$ becomes $-\frac{Z^2}{4 r_S^2}$, as a result of the fact that Flamm's coordinates are not asymptotically Lorentzian. 

The intrinsic geometry of Flamm's paraboloid and its isometric equatorial plane in the Schwarzschild spatial submetric differs critically from the intrinsic hyperbolic geometry of the Bolyai-Lobachevsky plane in at least two major respects. Firstly, the latter requires a constant negative intrinsic Gaussian curvature, whereas Flamm's paraboloid has $K = - \frac{r_S}{2} \frac{1}{r^3}$, rapidly vanishing for $r>>r_{S}$. At the surface of the earth, for example, we have $K \simeq - 1.7$ x 10$^{-27}$ cm$^{-2}$, far smaller than $K \simeq - 0.64$ cm$^{-2}$ at the Schwarzschild radius $r_{S} \simeq 0.887$ cm of a corresponding black hole.\cite{Rindler} Secondly, Flamm's paraboloid is a genus-one surface, excluding the $r < r_{S}$ hole region. That allows for the possibility of geodesic orbits encircling that hole region any number of times, and correspondingly crossing themselves. Thus, globally, infinitely many geodesics can possibly be drawn between any two points on the equatorial plane of the Schwarzschild spatial submetric, or, equivalently, on its isometric Flamm's paraboloid. 

The geodesic orbit \Eq{spatialradialgeodesicorbitperiastron} for $L \ne 0$ can be integrated by separation of variables as follows. Set $\hat{r} = r/r_{S}$ and $p = r_{p}/r_{S} >1$. The angle between two vectors with radii $\hat{r}_{1} \ge p$ and $\hat{r}_{2}>\hat{r}_{1}$ is thus
\begin{equation}\label{spatialsolutionintegral}
\phi(\hat{r}_{2},\hat{r}_{1}) = \int_{\hat{r}_{1}}^{\hat{r}_{2}} \frac{p \, \dd r}{\sqrt{r (r - 1) (r - p) (r + p)}}.
\end{equation} 
This integration can be performed either numerically or analytically. Numerical solutions require consideration of singularities in the integrand. An analytic solution is generally possible and more satisfactory, both theoretically and practically. It can be expressed in terms of elliptic integrals and functions as follows: 
\begin{widetext}
\begin{equation}\label{spatialradialgeodesicorbitperiastronsolution}
\phi(\hat{r}_{2},\hat{r}_{1}) = 2 \sqrt{\frac{p}{p-1}} \left( \textrm{F} \left[ \sin^{-1} \left( \sqrt{ \frac{(p - 1) (\hat{r}_{1} + p)}{2 p (\hat{r}_{1} - 1)} } \right) \bigg| \frac{-2}{p - 1}\right] -  \textrm{F} \left[ \sin^{-1} \left( \sqrt{ \frac{(p - 1) (\hat{r}_{2} + p)}{2 p (\hat{r}_{2} - 1)} } \right) \bigg| \frac{-2}{p - 1}\right] \right),
\end{equation} 
\end{widetext}
where $\textrm{F}\left[ \phi | m \right]$ is the incomplete elliptic integral of the first kind,
\begin{equation}
\textrm{F}\left[ \phi | m \right] = \int_{0}^{\phi} \left(1 - m \sin^2 \theta\right)^{-1/2} \, \dd \theta,
\end{equation}
for $-\pi/2 < \phi < \pi/2$. Extensions beyond this range of $\phi$ may be made using transformations of the argument as
\begin{equation}
\textrm{F}\left[n \pi \pm \phi| m \right] = 2 n \textrm{K}[m] \pm \textrm{F}\left[\phi | m \right],
\end{equation}
where $\textrm{K}\left[ m \right] = \textrm{F}\left[\pi/2 | m \right]$ is the complete elliptic integral of the first kind.\cite{Library} The general solution for $\phi(\hat{r})$, parameterized in terms of $p$, is given by
\begin{widetext}
\begin{align}\label{eqn:phiofrspatial}
\phi(\hat{r}) &= \lim_{\hat{r}_{1} \to p} \phi(\hat{r},\hat{r}_{1}) \nonumber\\
&=2 \sqrt{\frac{p}{p - 1}} \left( \textrm{K}\left[ \frac{-2}{p - 1} \right] - \textrm{F} \left[ \sin^{-1} \left( \sqrt{\frac{(p - 1) (p + \hat{r})}{2 p (\hat{r} - 1)}} \right) \bigg| \frac{- 2}{ p - 1 } \right] \right).
\end{align}
\end{widetext}
%
%
The asymptotic limit for the total angular deflection relative to the symmetry axis (let us say, the $X$-axis) is 
\begin{widetext}
\begin{align}\label{eqn:phiinfinity}
\phi_\infty &= \lim_{\hat{r} \to \infty} \phi(\hat{r})\nonumber\\
&=2 \sqrt{\frac{p}{p-1}} \left( \textrm{K} \left[ \frac{-2}{p-1}\right] - \textrm{F} \left[ \sin^{-1} \left( \sqrt{\frac{p-1}{2 p}}\right) \bigg|\frac{-2 }{p - 1}\right] \right).
\end{align}
\end{widetext}
Functional inversion of \Eq{eqn:phiofrspatial} uniquely provides an analytic solution to the spatial geodesic orbit \Eq{spatialradialgeodesicorbitperiastron}. That is
\begin{equation}\label{solutionspatialradialgeodesicorbitperiastron}
r(\phi) = \frac{p}{2 \, \text{cn}\left[ \sqrt{\frac{p-1}{4 p}} \phi
   \bigg|\frac{-2}{p-1}\right]^2-1} ,
\end{equation}
where $cn$ denotes the Jacobi elliptic cosine function, and the angle $\phi$ is taken within the range $(-\phi_{\infty}, \phi_{\infty})$.

Values for all these elliptic integrals and functions can be readily obtained from current computer packages. All the solutions that we illustrate in this paper, and many more for the same or other related metrics, have been derived from Mathematica libraries. At times, we checked analytic solutions with direct numerical integrations, confirming their accuracy.

\begin{figure}[!hb]
	\begin{center}
 	\includegraphics[width=8.6cm]{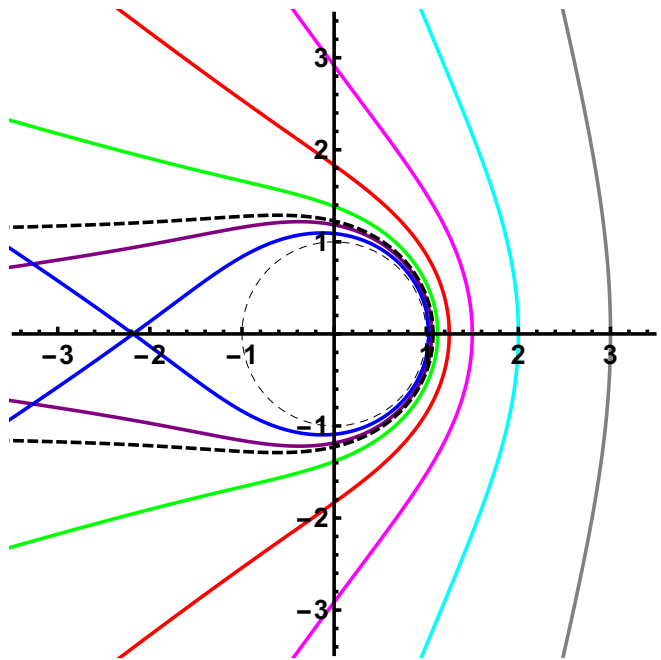}\\
      \caption{\label{fig:spatialgeodesics} Some spatial geodesics with periastra at $r_p/r_S$ = 3, 2, 1.5, 1.25, 1.125, 1.07611, 1.0625, 1.03125. Here $r_S = 1$. The critical $2\pi$-encircling geodesic orbit with $r_{p1} = 1.07611 r_S$ is displayed as a dashed line.} 
	\end{center}
\end{figure}

\begin{figure}[!hb]
	\begin{center}
 	\includegraphics[width=8.6cm]{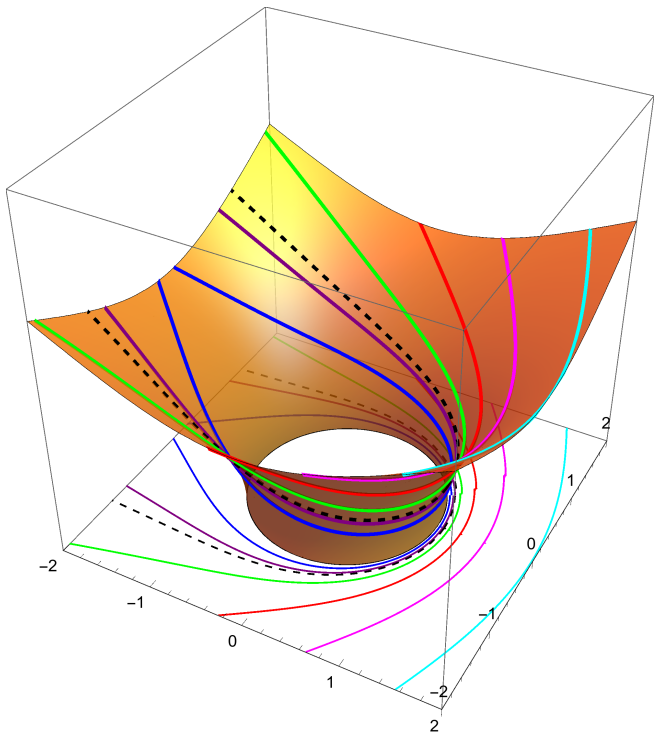}\\
      \caption{\label{fig:embeddedgeodesics} Geodesics on Flamm's paraboloid with periastra at $r_p/r_S$ = 2, 1.5, 1.25, 1.125, 1.07611, 1.0625, 1.03125, isometric to the geodesics shown in \Figref{fig:spatialgeodesics}. Here $r_S = 1$. The critical $2\pi$-encircling geodesic orbit with $r_{p1} = 1.07611 r_S$ is displayed as a dashed line.} 
	\end{center}
\end{figure}

As a first example, some geodesic orbits are graphed in \Figref{fig:spatialgeodesics} for $r_p/r_S = 3, 2, 1.5, 1.25, 1.125, 1.07611, 1.0625$, and 1.03125 on Schwarzschild equatorial plane. The corresponding geodesic orbits on the isometric Flamm's paraboloid are graphed in \Figref{fig:embeddedgeodesics}. Since we have \textit{azimuthal symmetry}, we have chosen without loss of generality to align the $X$-axis along a direction from $r_S$ to $r_p$ in \Figref{fig:spatialgeodesics} and \Figref{fig:embeddedgeodesics}. 

Notice that $L \ne 0$ geodesics on Flamm's paraboloid, as illustrated in \Figref{fig:embeddedgeodesics} for example, have nothing to do with \textit{circles} at $Z_0$ = const $\ne 0$ heights, as typically drawn on similar figures, such as that displayed on the cover of Hartle's book, for example.\cite{Hartle} Those circles are level curves with $\hat{r}_{0}$ = const $>$ 1 and non-zero geodesic curvature $\kappa_g = \frac{1}{r_S} \sqrt{ \frac{(\hat{r}_{0} - 1)}{\hat{r}_{0}^3}}$. See Refs.~\onlinecite{Resca, O'Neill} for further discussions on that matter.

Remarkably, even in the non-relativistic limit of $r_p >> r_S$, the short-range relativistic attractive fourth term in \Eq{spatialradialgeodesicorbitperiastron} still affects the asymptotic behavior of nearly Euclidean straight lines, causing their semi-asymptotes to form a \textit{concave} angle $2\phi_\infty$ slightly larger than $\pi$, about the $X$-axis in \Figref{fig:spatialgeodesics} for example. That differs from the hyperbola solution of the three-term non-relativistic approximation to \Eq{spatialradialgeodesicorbitperiastron}, whose semi-asymptotes form a \textit{convex} angle slightly smaller than $\pi$. Nevertheless, the effect of the long-range non-relativistic repulsive second term in the full \Eq{spatialradialgeodesicorbitperiastron} is always noticeable as an asymptotic bending \textit{away} from the hole region for \textit{all} geodesics: observe those in \Figref{fig:spatialgeodesics}, for example. Indeed, the second term always exceeds in magnitude the fourth term in \Eq{spatialradialgeodesicorbitperiastron}, except at $r = r_p$, where they equilibrate.

The onset of a fully relativistic regime and multiple connectivity can be characterized by a critical periastron $r_{p1}= 1.07611 r_S$, where we attain the first full encircling of the $r<r_S$ hole region, but without any crossing of the geodesic orbit. Such $r_{p1}$ is determined by solving for $\phi_\infty = \pi$ in \Eq{eqn:phiinfinity}. That produces a full \textit{concave} angle $2\phi_\infty$ infinitesimally smaller than $2\pi$ between geodetic semi-asymptotes. Below that $r_{p1}$ value, there is an infinite series of $r_{pn}$ periastra that decreasingly converge to $r_S$, such that their corresponding geodesics have increasing integer numbers of windings around the hole region and corresponding crossings. The critical $2\pi$-encircling geodesic orbit with $r_{p1}= 1.07611 \, r_S$ is displayed as a dashed line in \Figref{fig:spatialgeodesics} and in \Figref{fig:embeddedgeodesics}. 

\begin{figure}[!hb]
	\begin{center}
   	\includegraphics[width=8.6cm]{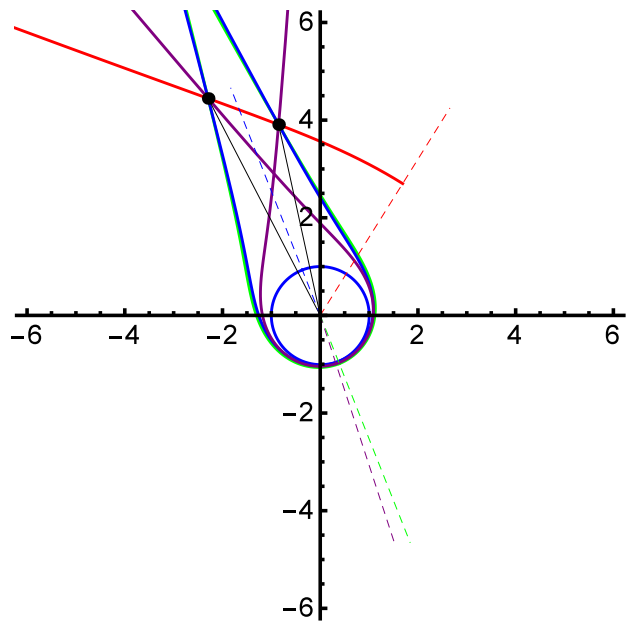}\\
      \caption{\label{fig:multiplegeodesics1} Four spatial geodesics connecting two points on the equatorial plane of Schwarzschild spatial geometry, where $r_S$ is taken as the unit of length. One point has $r_1 = 4 r_S$ and $\phi_1 = \pi/4 + 1$, while the other point has $r_2 = 5 r_S$ and $\phi_2 = \pi/3 + 1$. Dashed radial rays indicate the positive side of symmetry axes and the location of periastra of correspondingly colored geodesics. Periastra are at $r_p/r_S = 3.1838$ (red), 1.06209 (green), 1.04152 (purple), 1.00065 (blue).}
	\end{center}
\end{figure}

Typically, we can pick two points on the equatorial geodesic plane and find an infinite number of longer and longer arcs of geodesics that connect them, spiraling in-and-out around the hole region. In \Figref{fig:multiplegeodesics1} we provide some examples of that. One point has $r_1 = 4 r_S$ and $\phi_1 = \pi/4 + 1$, while the other point has $r_2 = 5 r_S$ and $\phi_2 = \pi/3 + 1$. Periastra occur at $r_p/r_S = 3.1838$ (red), 1.06209 (green), 1.04152 (purple), 1.00065 (blue). All periastra fall on the positive side of symmetry axes that are shown as dashed rays from the origin. The red curve provides the shortest geodesic arc between the two points. The red geodesic never crosses itself nor fully encircles the hole region. The green geodesic encircles the hole region once, crossing itself at a point on the negative side of the symmetry axis. The purple geodesic still encircles the hole region once, and still crosses itself only once on the negative side of the symmetry axis. However, the crossing point of the purple geodesic now falls closer to the hole region, on the same side of the red geodesic. The blue geodesic finally encircles the hole region twice, crossing itself at two points on both sides of the symmetry axis.

\begin{figure}[!hb]
	\begin{center}
 	\includegraphics[width=8.4cm]{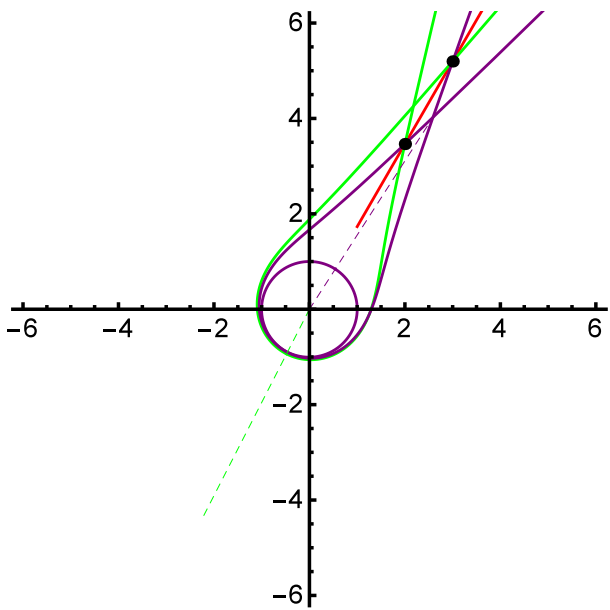}\\
      \caption{\label{fig:multiplegeodesics2} Three spatial geodesics connecting two points on the equatorial plane of Schwarzschild spatial geometry, where $r_S$ is taken as the unit of length. The connected two points lie along the same radial direction, having $\phi = \pi/3$. One point has $r_1 = 4 r_S$, while the other point has $r_2 = 6 r_S$. Dashed radial rays indicate the positive side of symmetry axes and the location of periastra of correspondingly colored geodesics. Periastra are at $r_p/r_S = 1$ (red), 1.05228 (green), 1.00056 (purple).}
	\end{center}
\end{figure}

In \Figref{fig:multiplegeodesics2} two points are taken along the same radial direction, having $\phi = \pi/3$. One point has $r_1 = 4 r_S$, while the other point has $r_2 = 6 r_S$. Periastra are at $r_p/r_S = 1$ (red), 1.05228 (green), 1.00056 (purple). The red segment provides the shortest radial ($L=0$) connection between the two points. The green geodesic encircles the hole region once, crossing itself on the negative side of the symmetry axis. The purple geodesic encircles the hole region twice and crosses itself twice, on both sides of the symmetry axis.

To make further progress, we need to reframe the spatial geodetic analysis in terms of impact parameters. Given the fact that the intrinsic Gaussian curvature, $K = - \frac{r_S}{2} \frac{1}{r^3}$, vanishes asymptotically, we may pursue the analogy with the SR asymptotic limit of GR Schwarzschild space-time for $r \rightarrow \infty $. Then the $L$ angular momentum equivalent defines the impact parameter equivalent as 
\begin{equation}\label{impactparameter}
b^2 = {\frac{L^2}{C^2}} .
\end{equation}
Considering \Eq{periastron}, we find that $b = r_p$. Geometrically, $b$ represents the distance between either incoming or outgoing geodetic asymptotes and corresponding radial lines (with $L=b=0$) asymptotically parallel to those geodetic asymptotes.  

\begin{figure}[!hb]
	\begin{center}
 	\includegraphics[width=8.5cm]{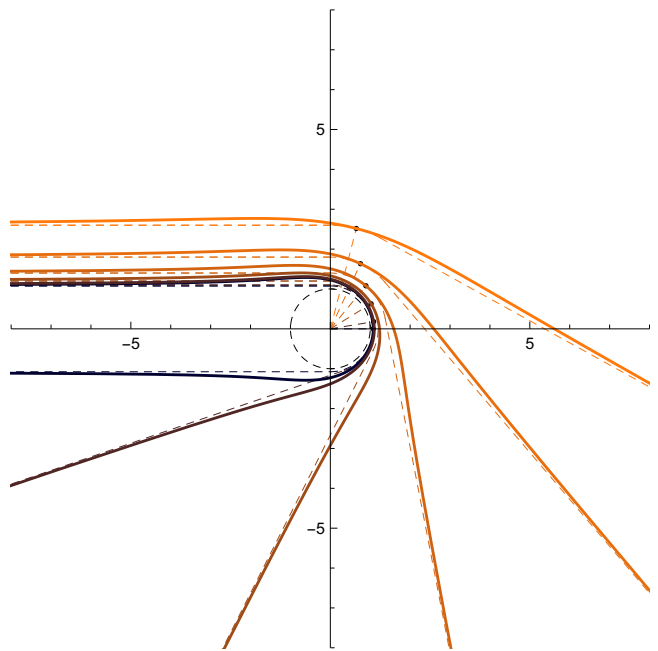}\\
      \caption{\label{fig:impact} Solid curves are spatial geodesics of varying impact parameters ($r_p/r_S = 1.07611, 1.1, 1.2, 1.4, 1.8,$ and 2.6). The corresponding dashed curves match smoothly asymptotic lines with arcs of constant radius. The points and radial dashed lines identify the point of closest approach and the symmetry axis, respectively.}
	\end{center}
\end{figure}

In \Figref{fig:impact} we display geodesics with varying impact parameters, starting with the critical $2\pi$-encircling geodesic orbit with $r_{p1}/r_S = 1.07611$, and continuing with $r_p/r_S = 1.1, 1.2, 1.4, 1.8, 2.6$. The corresponding dashed curves match  asymptotic lines smoothly with arcs of constant radius. If we further decrease $b$ from $r_{p1}$ to $r_S$, the geodesic orbit will cross itself an increasing number of times, approaching infinity for $b$ approaching $r_S$. 

\section{Singular spatial geodesics}\label{sec:s-geodesics}

There is a different class of singular geodetic solutions, which we may call $s$-geodesics. Those occur for impact parameters having $b_a < r_S$, thus $p_a=b_a/ r_S < 1$. Periastra of $s$-geodesics only occur at $r_S$. Impact parameters $b_a$, although still given by \Eq{impactparameter}, actually correspond to apoastra of geodesics in the hole region with indefinite metric in \Eq{spatialmetric}, which we shall not further discuss in this paper. Suffice it to say that no spatial geodesic can cross the horizon at $r=r_S$ from one region to the other, whether out-going or in-coming at any angle. Of course that represents yet another major difference with space-time geodesics, which can definitely cross the horizon at $r=r_S$.

In order to understand the singularity of $s$-geodesics at $r_S$, we must return to \Eq{spatialradialgeodesic}, which in fact represents a first-integral. Taking $\frac{d}{d\lambda}$ of that, or, equivalently, working out the standard form of the geodesic equation for contravariant components with Christoffel symbols, we obtain

\begin{equation}\label{Christoffel}
\frac{d^2r}{d\lambda^2}  = \frac{r_S}{2r^4} (C^2 r^2 - L^2) + (r - r_S)  \frac{L^2}{r^4} .
\end{equation}

For regular geodesics, having $b>r_S$, evaluation of \Eq{Christoffel} at their $r_p$ periastron yields
\begin{equation}\label{r-geodesic}
  \bigg(\frac{d^2r}{d\lambda^2}\bigg)_{r_p} = (r_p -r_S)  \frac{L^2}{r_p^4} > 0 .
\end{equation}

However, for $s$-geodesics having $b_a < r_S$, evaluation of \Eq{Christoffel} at their $r_S$ periastron yields

\begin{equation}\label{s-geodesic}
 \bigg(\frac{d^2r}{d\lambda^2}\bigg)_{r_S} =  \frac{L^2}{2r_S} \bigg(\frac{1}{b_a^2} - \frac{1}{r_S^2} \bigg)  > 0 .
\end{equation}

\begin{figure}[!hb]
	\begin{center}
 	\includegraphics[width=8.6cm]{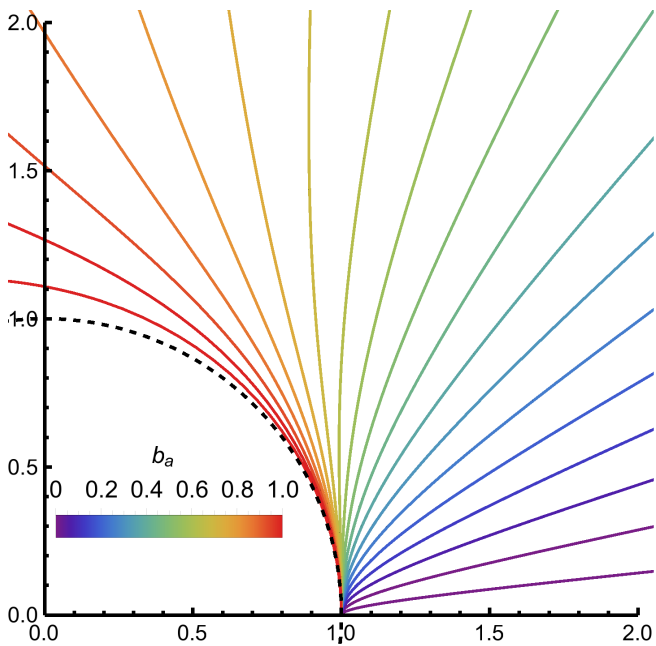}\\
      \caption{\label{fig:array} Infinite number of $s$-geodesics, all tangent to each other and to the $r=r_S=1$ circle at $\phi = 0$, transporting the same initial tangent vector parallel to itself and yet in all subsequently different directions.}
	\end{center}
\end{figure}

Clearly, the `acceleration equivalent' in \Eq{r-geodesic} has a single value, whereas \Eq{s-geodesic} involves a continuous range of possibilities, having $0 < b_a < r_S$. Thus, starting at any point with $r \ge r_p > r_S$ with any initial vector, there is a unique regular geodesic that transports that vector parallel to itself indefinitely. However, starting at any point on the horizon, where $g_{rr}$ diverges, there is an infinite number of $s$-geodesics, all tangent to each other and to the $r=r_S$ circle, that transport the same initial tangent vector parallel to itself and yet in all subsequently different directions. This situation is depicted in \Figref{fig:array}. 

The situation is regularized on Flamm's paraboloid, if we consider both surfaces with positive and negative $Z$-values, joined at the $Z=0$ circle. In that perspective, the $(Z,\phi)$ coordinates produce a line element 
\begin{equation}\label{reducedspatialmetricFlamm}
  d S^2= \bigg(1 + \frac{Z^2}{4r_S^2} \bigg) (dZ)^2 + r_S^2 \bigg(1 + \frac{Z^2}{4r_S^2} \bigg)^2 (d \phi)^2  .
\end{equation} 

That element is equal in value to $d S^2$ in \Eq{reducedspatialmetric} for the $(r,\phi)$ coordinates, but the $Z=0$ circle is no longer represented as a line of coordinate singularities in \Eq{reducedspatialmetricFlamm}. 

Thus, in \Eq{spatialradialgeodesicorbitperiastron} with $b_a$ replacing $r_p$, the $Z$-elevation of $s$-geodesics produces a unique tangent vector that intersects the $Z=0$ circle at a specific angle $\gamma$ such that 

\begin{equation}\label{s-geodesic-angle}
\tan(\gamma) = \bigg| \bigg(\frac{dZ}{r d\phi}\bigg)_{r_S} \bigg| = \sqrt{\frac{r_S^2}{b_a^2} - 1} .
\end{equation}

\begin{figure}[!hb]
	\begin{center}
 	\includegraphics[width=8.6cm]{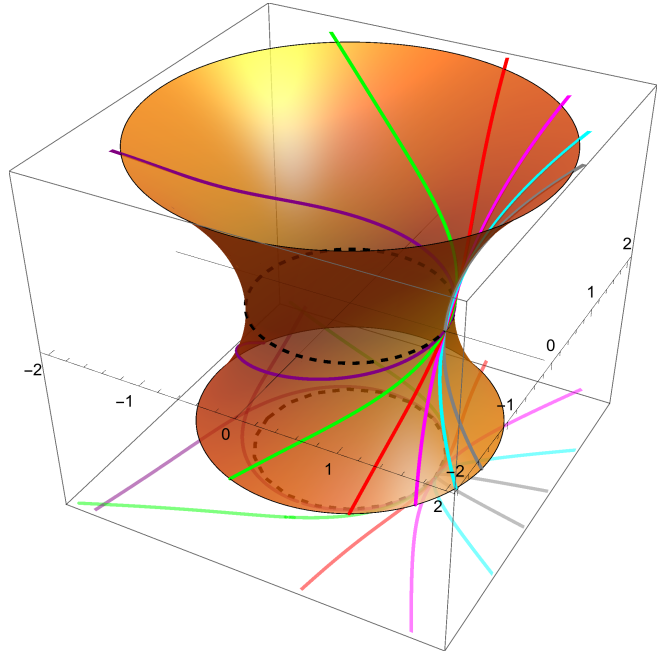}\\
      \caption{\label{fig:embeddedgeosb} Solid curves are $s$-geodesics of varying $p_a$ (0.1, 0.3, 0.5, 0.7, 0.9, and 0.99). The dashed circle indicates a radius of $r_S = 1$. Notice that only the innermost purple $s$-geodesic with the greater $p_a =0.99$ fully encircles the hole region once.}
	\end{center}
\end{figure}

Examples of that are shown in \Figref{fig:embeddedgeosb}. 

Regularization of $s$-geodesics on the full Flamm's paraboloid can also be appreciated by considering the relation

\begin{equation}\label{Z-r-relation}
  \bigg(\frac{dZ}{d\lambda}\bigg)^2
    = \frac{r_S}{r} g_{rr} \bigg(\frac{dr}{d\lambda}\bigg)^2  .
\end{equation}
That produces a well defined limit

\begin{equation}\label{Z-relation-limit}
 \bigg(\frac{dZ}{d\lambda}\bigg)^2_{r_S} =  \frac{L^2}{b_a^2} - \frac{L^2}{r_S^2}  > 0 
\end{equation}
for $r \rightarrow r_S$, yielding \Eq{s-geodesic-angle}, even though 

\begin{equation}\label{r-relation-limit}
 \bigg(\frac{dr}{d\lambda}\bigg)^2_{r_S} =  0.
\end{equation}

In terms of $(r,\phi)$ coordinates, $s$-geodesic solutions are obtained from \Eq{spatialsolutionintegral} by setting its lower limit at $\hat{r}_1 = 1$ and by replacing $p>1$ with $p_a < 1$. Thus we obtain

\begin{widetext}
\begin{align}\label{eqn:phiofrspatialothers}
\phi_s(\hat{r}) &=  \int_{1}^{\hat{r}} \frac{p_a \, \dd r}{\sqrt{r (r - 1) (r - p_a) (r + p_a)}} \nonumber\\
&=2 \sqrt{\frac{p_a}{1-p_a}} \left( \textrm{K}\left[- \frac{1+p_a}{1 - p_a} \right] + i \textrm{F} \left[ i \sinh^{-1} \left( \sqrt{\frac{(1 - p_a) (p_a + \hat{r})}{2 p_a (\hat{r} - 1)}} \right) \bigg| \frac{2}{ 1 - p_a } \right] \right).
\end{align}
\end{widetext}
In the limit of $r \rightarrow \infty$, the final angle is given by
\begin{widetext}
\begin{equation}
\phi_{s,\infty} = \lim_{\hat{r} \to \infty} \phi_s(\hat{r}) = 2 \sqrt{\frac{p_a}{1 - p_a}}
   \left(\textrm{K} \left[- \frac{1+p_a}{1-p_a}\right ]+i \textrm{F} \left[i \sinh  ^{-1}\left(\sqrt{\frac{1-p_a}{2 p_a}}\right)\bigg|\frac{2}{1-p_a}\right]\right).
\end{equation}
\end{widetext}
We may also invert \Eq{eqn:phiofrspatialothers} to obtain an explicit expression of $\hat{r}_s$ as a function of $\phi$,
\begin{equation}
\hat{r}_s (\phi) = \frac{p_a \left(1 - \textrm{sn} \left[ i \sqrt{\frac{1-p_a}{4 p_a}}, \frac{2}{1-p_a} \right]^2 \right)}{\textrm{sn} \left[ i \sqrt{\frac{1-p_a}{4 p_a}}, \frac{2}{1-p_a} \right]^2+p_a},
\end{equation}
where $sn$ denotes the Jacobi elliptic sine function, and the angle $\phi$ is taken within the range $(-\phi_{s,\infty}, \phi_{s,\infty})$. 

\begin{figure}[!hb]
	\begin{center}
 	\includegraphics[width=8.6cm]{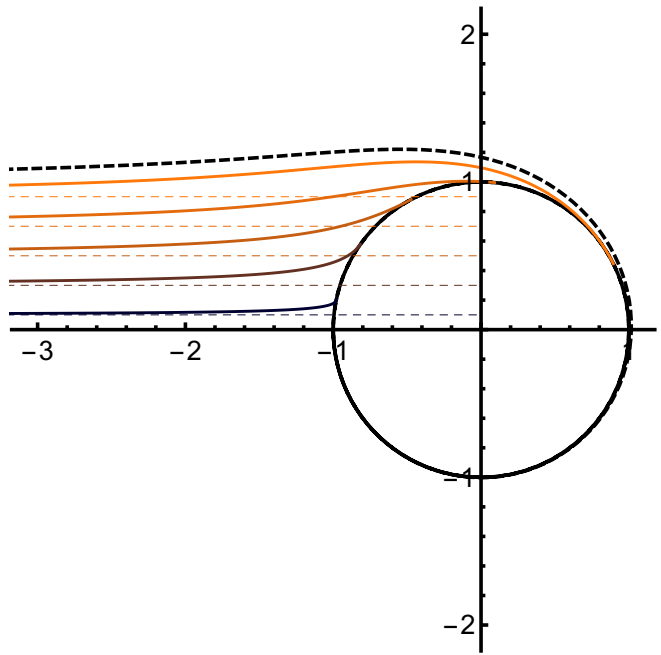}\\
      \caption{\label{fig:impactb} Solid curves are $s$-geodesics of varying impact parameters ($p_a = 0.1, 0.3, 0.5, 0.7,$ and 0.9). The corresponding dashed lines represent their asymptotes, parallel to the negative $X$-axis. The dashed black line corresponds to the $p_a = 1$ limiting case.}
	\end{center}
\end{figure}

Examples of $s$-geodesics of varying impact parameters, with $p_a = b_a/r_S \le 1$, are shown in \Figref{fig:impactb}. Notice again that all $s$-geodesics obey \Eq{r-relation-limit}, i.e., the $r_S$-tangential condition. Continuation of $s$-geodesics at the horizon is not shown in \Figref{fig:array} or \Figref{fig:impactb}, while it is shown through the full Flamm's paraboloid in \Figref{fig:embeddedgeosb}. 

Having shown that $s$-geodesics parallel-transport their tangent vectors continuously above and below the $Z=0$ circle on the full Flamm's paraboloid, it is best to isometrically view the Schwarzschild geodesic equatorial plane as having two sides, joined at the horizon. Therein, $s$-geodesics parallel-transport their tangent vectors continuously through the $r=r_S$ horizon from the upper to the lower side, or conversely.

For $b_a = 0$ there are only radial geodesics, derived from \Eq{spatialradialgeodesic} for $L = 0$. That corresponds to continuous parabolae spanning both positive and negative $Z$-values on the full Flamm's paraboloid. Those parabolae intersect vertically the $Z=0$ circle, with an angle $\gamma = \pi/2$, according to \Eq{s-geodesic-angle}.

For the critically separating value of $b = b_a = r_S$, the point-particle spirals around the $r_S$ circle infinitely many times, without ever reaching it exactly. If it did, the geodesic would transform into that of the $r = r_S$ circle. The angle $\gamma$ in \Eq{s-geodesic-angle} vanishes in that limit. It is still possible to solve analytically \Eq{spatialsolutionintegral} for $p=p_a=1$, obtaining
\begin{widetext}
\begin{equation}\label{limitingintegral}
\phi_c(\hat{r}) = \int_{\hat{r}}^{\infty} \frac{\dd z}{(z - 1) \sqrt{z (z + 1)}} = \frac{1}{\sqrt{2}} \ln \left(\frac{\left(3-2
   \sqrt{2}\right)
   \left(\sqrt{\hat{r}}+1\right)
   \left(\sqrt{\hat{r}}+\sqrt{2}
   \sqrt{\hat{r}+1}+1\right)}{\left(\sqrt{\hat{r}
   }-1\right)
   \left(-\sqrt{\hat{r}}+\sqrt{2}
   \sqrt{\hat{r}+1}+1\right)}\right).
\end{equation} 
\end{widetext}
This solution is plotted as the dashed black curve in \Figref{fig:impactb}, asymptotically starting parallel to the negative $X$-axis with $p_a = 1$ impact separation.

\begin{figure}[!hb]
	\begin{center}
 	\includegraphics[width=8.6cm]{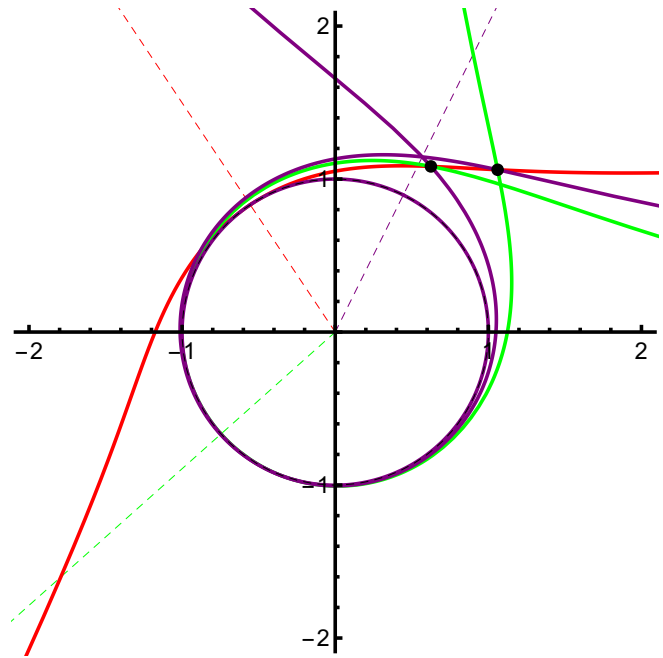}\\
      \caption{\label{fig:multiplegeodesicswithnewgeos} Three $s$-geodesics connecting two points of the equatorial plane of Schwarzschild spatial geometry. One point has $r_1 = 1.5 r_S$ and $\phi_1 = \pi/4$, while the other point has $r_2 = 1.25 r_S$ and $\phi_2 = \pi/3$. Dashed radial rays indicate the positive side of symmetry axes and the location of periastra of correspondingly colored geodesics. All periastra occur at $r_S =1$. Impact parameters are $b_a/r_S = 0.807342$ (red), 0.98445 (green), 0.999873 (purple). Only the red $s$-geodesic connects two points on the same side of the Flamm embedding.  The green and purple geodesics connect two points on opposite sides of the the Flamm embedding. Green and purple $s$-geodesics fully encircle the hole region once and twice, respectively, while the red $s$-geodesic never does.}
	\end{center}
\end{figure}

\begin{figure}[!hb]
	\begin{center}
 	\includegraphics[width=8.6cm]{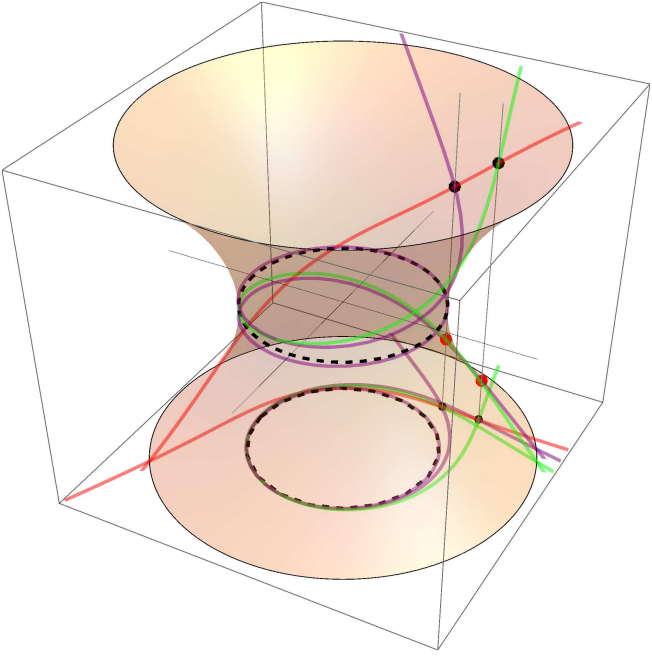}\\
      \caption{\label{fig:multiplegeodesicswithnewgeosFlamm} Flamm embedding of $s$-geodesics connecting two points on either sides of the equatorial plane in \Figref{fig:multiplegeodesicswithnewgeos}.}
	\end{center}
\end{figure}

Let us now consider again any two points on the top side of the equatorial geodesic plane, say, or on the top surface of Flamm's paraboloid, equivalently. It is not always possible to directly connect these two points with regular geodesics, as we did in \Figref{fig:multiplegeodesics1} and \Figref{fig:multiplegeodesics2}, for example. When regular geodesics cannot directly connect the two points, $s$-geodesics can, and vice versa. There are differences, however. Only the shortest $s$-geodesic arc truly connects the two points on the same side of the equatorial geodesic plane. Longer $s$-geodesic arcs that may or may not encircle the hole region any number of times are bound to fall into the other side of the equatorial geodesic plane. From the perspective of the full Flamm's paraboloid, longer $s$-geodesic arcs thus only connect points having opposite signs in their $Z$-coordinates. Examples of this behavior are shown in \Figref{fig:multiplegeodesicswithnewgeos} on the equatorial geodesic plane, and more clearly in \Figref{fig:multiplegeodesicswithnewgeosFlamm} on its equivalent full Flamm's paraboloid. Conversely, regular geodesics are bound to one side of the equatorial geodesic plane. Thus, regular geodesics cannot connect points having opposite signs in their $Z$-coordinates on the corresponding full Flamm's paraboloid.


\begin{figure}[!hb]
	\begin{center}
 	\includegraphics[width=8.6cm]{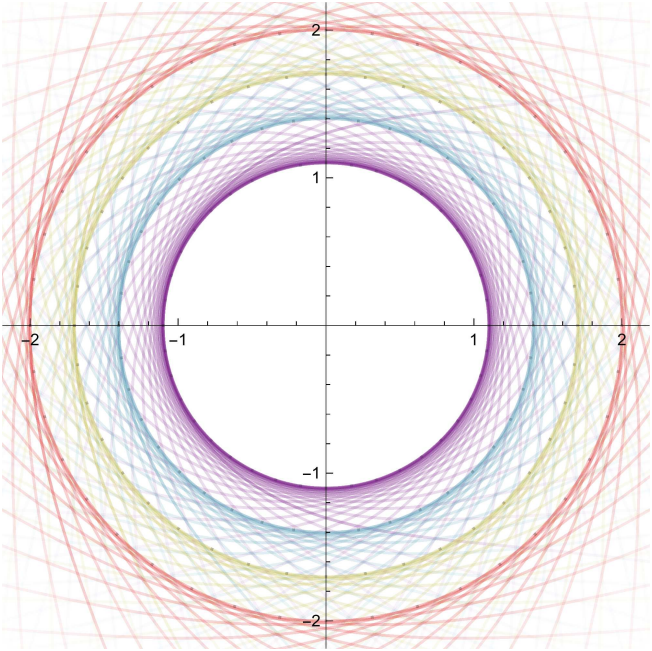}\\
      \caption{\label{fig:allold} Four grids of regular geodesics spanning each side of the spatial equatorial plane. Periastra originating each grid are taken at $r_p/r_S=$ 1.1, 1.4, 1.7, and 2.0.}
	\end{center}
\end{figure}

\begin{figure}[!hb]
	\begin{center}
 	\includegraphics[width=8.6cm]{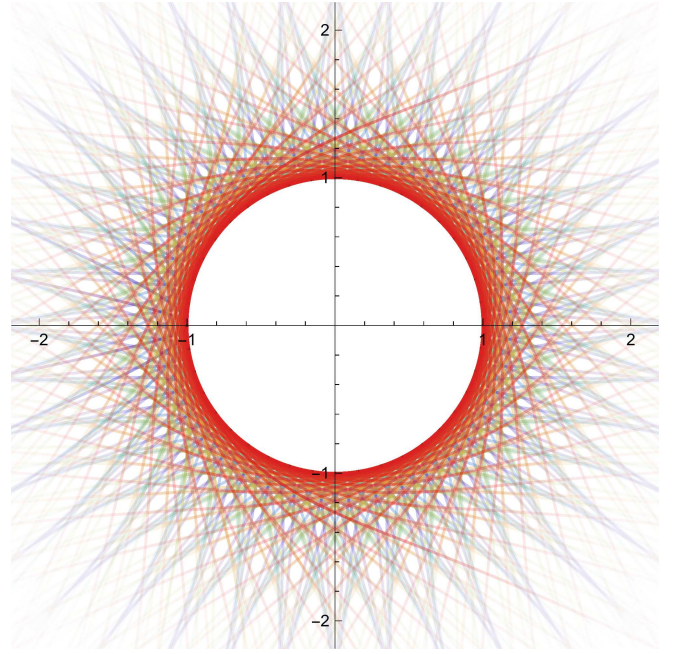}\\
      \caption{\label{fig:allnew} Nine grids of $s$-geodesics spanning both sides of the spatial equatorial plane. All periastra occur at $r_S = 1$, with impact parameters $b_a/r_S$ ranging from 0.1 to 0.9 in steps of 0.1.}
	\end{center}
\end{figure}

\begin{figure}[!hb]
	\begin{center}
 	\includegraphics[width=8.6cm]{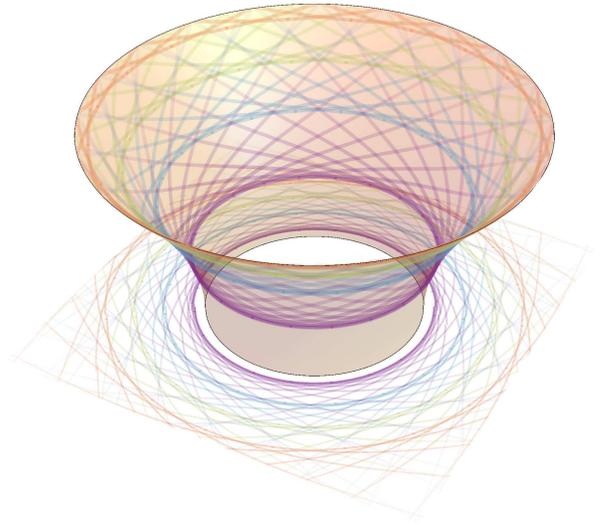}\\
      \caption{\label{fig:alloldFlamm} Flamm embedding of four grids of regular geodesics, spanning each side of the spatial equatorial plane in \Figref{fig:allold}.}
	\end{center}
\end{figure}

\begin{figure}[!hb]
	\begin{center}
 	\includegraphics[width=8.6cm]{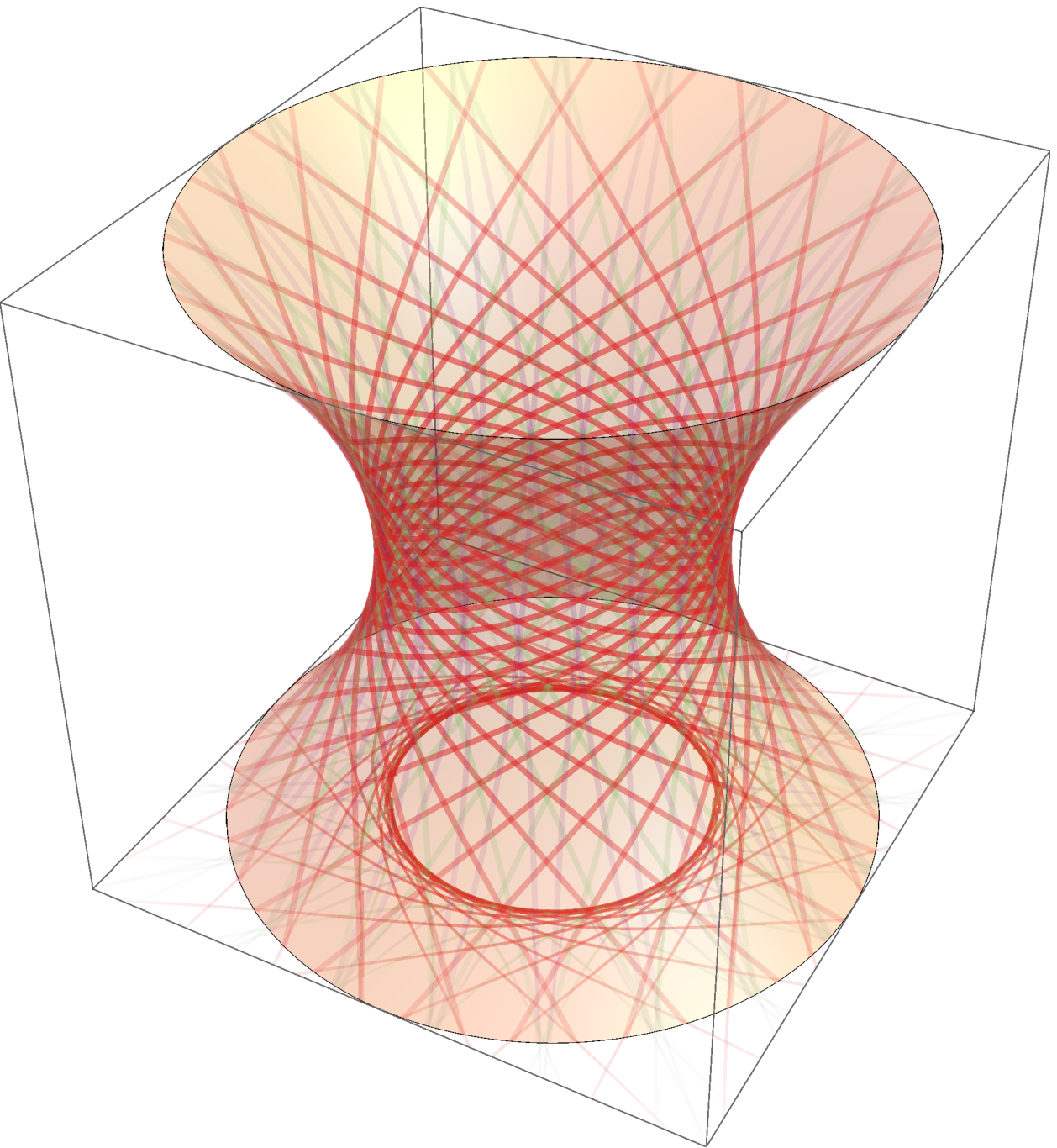}\\
      \caption{\label{fig:allnewFlamm} Flamm embedding of three grids of $s$-geodesics, spanning both sides of the spatial equatorial plane in \Figref{fig:allnew}. All periastra occur at $r_S = 1$. Impact parameters are $b_a/r_S = $ 0.1, 0.5, and 0.9.}
	\end{center}
\end{figure}

Regular and $s$-geodesics together provide geodesic completeness, forming a one-parameter family of curves with impact parameters ranging from $-\infty$ to $+\infty$. Further adding azimuthal symmetry, we may get a sense of the structure and space-filling distribution of regular and $s$-geodesics on the equatorial geodesic plane from \Figref{fig:allold} and \Figref{fig:allnew}, respectively. Equivalent renditions of their embeddings on half and full Flamm's paraboloids are shown in \Figref{fig:alloldFlamm} and \Figref{fig:allnewFlamm}, respectively. 


\begin{figure}[!hb]
	\begin{center}
	\includegraphics[width=8.6cm]{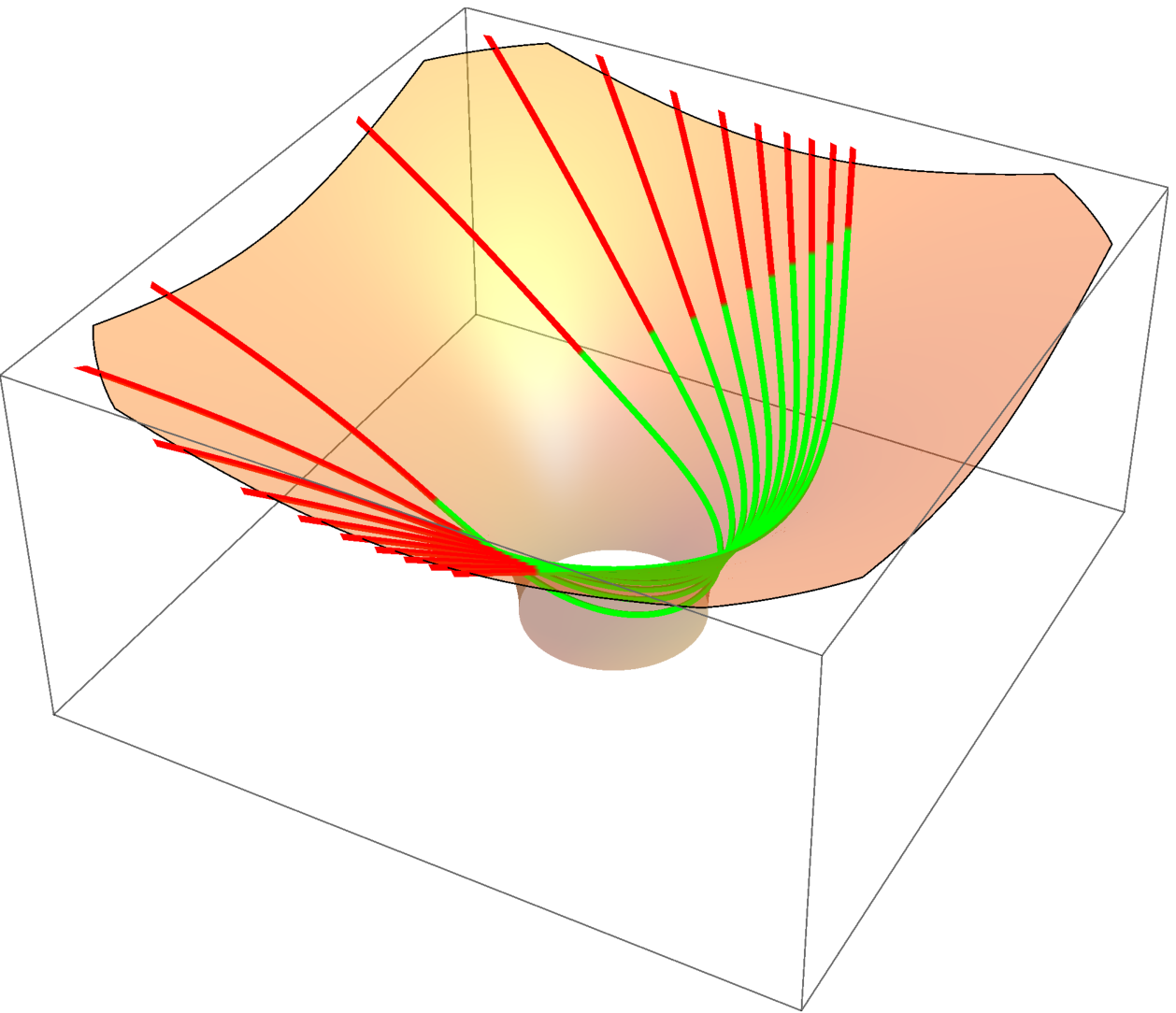}\\
    \caption{\label{fig:regularnormalcurvatureFlamm} Arcs of regular geodesics with positive (negative) normal curvatures are shown in green (red) on Flamm's paraboloid. }
	\end{center}
\end{figure}

\begin{figure}[!hb]
	\begin{center}
	\includegraphics[width=8.6cm]{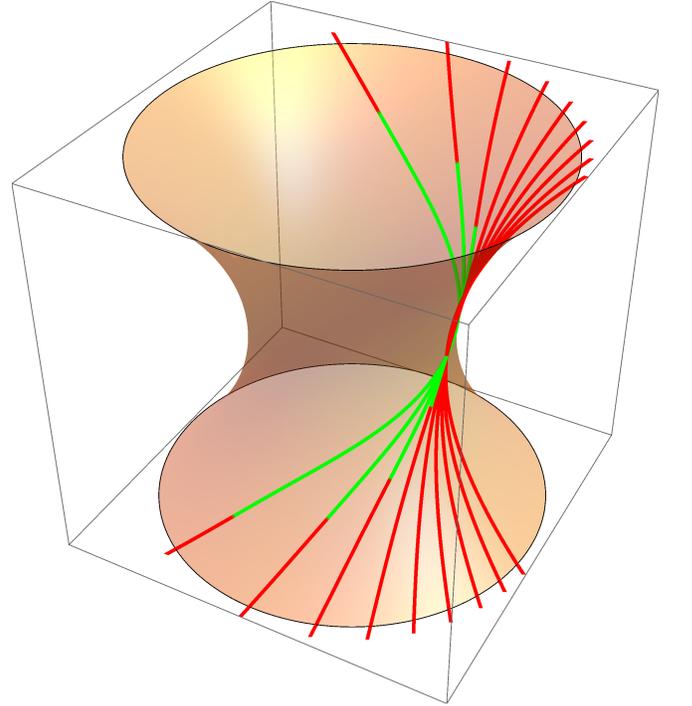}\\
     \caption{\label{fig:singularnormalcurvatureFlamm} Arcs of $s$-geodesics with positive (negative) normal curvatures are shown in green (red) on Flamm's paraboloid. The (normal) curvature of radial ($L=0$) geodesics, $\kappa_n = \kappa = - \frac{1}{2} \sqrt{ \frac{r_S}{r^3}}$, is always negative, pointing away from the hole region, although vanishing asymptotically.}
	\end{center}
\end{figure}

\begin{figure}[!hb]
	\begin{center}
	\includegraphics[width=8.6cm]{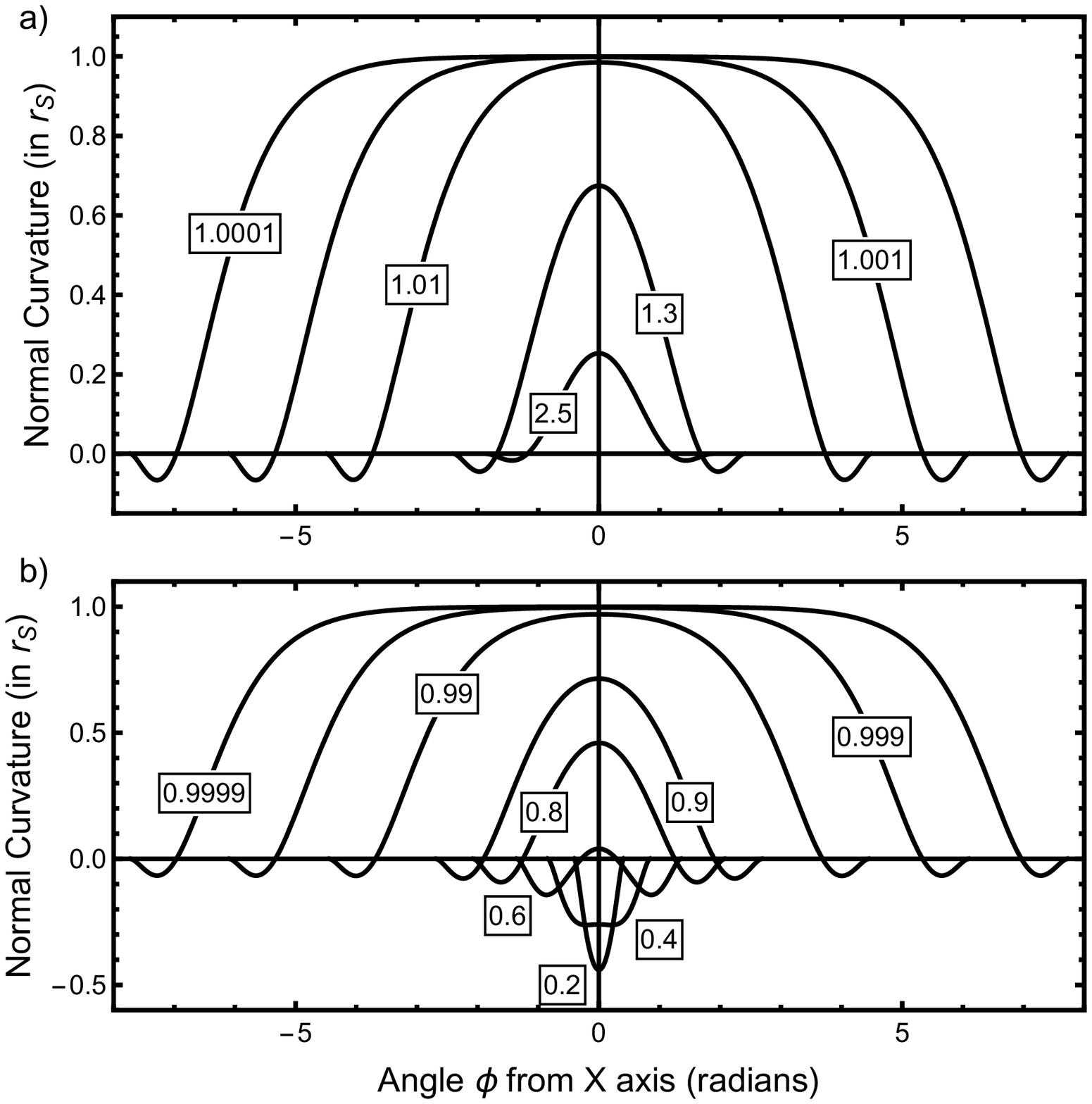}\\
      \caption{\label{fig:normalcurvatures} Plots of normal curvature  $\kappa_n$ (in units of $1/r_S =1$) vs. azimuthal angle $\phi$ (in radians) from the symmetry $X$-axis for: (a) five regular geodesics labeled by their periastra $r_p/r_S > 1$ within rectangular boxes; (b) eight $s$-geodesics labeled by their impact parameters $b_a/r_S < 1$ within rectangular boxes.}
	\end{center}
\end{figure}

It is of further interest to study independently geodesic curvatures, $\kappa_g$, normal curvatures, $\kappa_n$, and relative torsions, $\tau_r$, of curves embedded on Flamm's paraboloid, using standard notions and elements of differential geometry.\cite{do Carmo, Pressley} Geodesic curvatures must of course vanish for all geodesics. Normal curvatures are illustrated on Flamm's paraboloid in \Figref{fig:regularnormalcurvatureFlamm} for regular geodesics and in \Figref{fig:singularnormalcurvatureFlamm} for $s$-geodesics, respectively. Plots of corresponding normal curvatures are shown in \Figref{fig:normalcurvatures}. Loci of vanishing normal (thus total) curvatures reflect the varying hyperbolic geometry of Flamm's paraboloid. We have derived analytically and verified numerically many other differential form and curvature results. Ultimately, however, all that analysis and results can be obtained from the central geodesic orbit \Eq{spatialradialgeodesicorbitperiastron} and its exact solutions that we have already provided. Therefore, we shall not further report on such a complementary line of inquiry within this context.


\section{Geodesic orbits in curved time}\label{curvedtime}
Let us alternatively consider a 4D pseudo-Riemannian manifold with metric 
\begin{align}\label{metricwithoutspatialcurvature}
ds^2  = & g_{\mu \nu} dx^{\mu} dx^{\nu} \nonumber\\
      = & -\bigg( 1-\frac{r_S}{r} \bigg)(cdt)^2 + (dr)^2 + \nonumber\\
        & r^2 (d \theta)^2 + r^2 \sin^2 \theta (d \phi)^2.
\end{align}
This differs from the physically correct Schwarzschild metric in that the 3D spatial submanifold at any given coordinate-time, $t$, is devoid of any curvature in \Eq{metricwithoutspatialcurvature}. 

Following the same procedures that we adopted earlier produces now the \textit{time-like} geodesic orbit equation 
\begin{equation}\label{exactnospatialcurvatureradialgeodesicorbit}
  \bigg(\frac{d r}{d \phi}\bigg)^2
    = \frac{r^4}{\tilde{L}^2} \bigg\{c^2\tilde{E}^2 \bigg(1 - \frac{G}{c^2} \frac{2M}{r} \bigg)^{-1} - c^2  - \frac{\tilde{L}^2}{r^2}  \bigg\}.
\end{equation}
This agrees with the Schwarzschild result in the non-relativistic Newtonian limit, whereas geodesic orbit equations for the previous `splittable space-time' metric, \Eq{metricwithouttimecurvature}, do \textit{not}.\cite{Resca} Historically, that was instrumental for Einstein to realize that Newtonian gravity basically derives from the equivalence principle and its association with the gravitational redshift, even without full knowledge of Einstein field equations: cf. Chap. 18 of Ref. \onlinecite{SchutzGravity}, for example.

However, the \textit{null} geodesic orbit equation for the gravitational red-shift or `curved-time' metric of \Eq{metricwithoutspatialcurvature} is
\begin{equation}\label{exactnospatialcurvatureradialnullgeodesicorbit}
  \bigg(\frac{d r}{d \phi}\bigg)^2
    = \frac{r^4}{L^2} \bigg\{\frac{E^2}{c^2} \bigg(1 - \frac{G}{c^2} \frac{2M}{r} \bigg)^{-1} - \frac{L^2}{r^2}  \bigg\},
\end{equation}
which differs profoundly from the exact \textit{null} geodesic orbit \Eq{nullgeodesicorbitradial} of Schwarzschild space-time metric. 

Remarkably, however, turning points or apsides for both \textit{time-like} and \textit{null} geodesics coincide for both exact and `curved-time' metrics, \Eq{metric} and \Eq{metricwithoutspatialcurvature}. In particular, apsides of \textit{null} geodesics for both metrics satisfy the same cubic equation
\begin{equation}\label{periastra}
\alpha p^3 = p - 1,
\end{equation} 
where 
\begin{equation}\label{eqn:alpha}
\alpha = \frac{r_S^2 E^2}{c^2 L^2}.
\end{equation}

The three algebraic solutions to the cubic \Eq{periastra} add up to zero, according to Vieta's formula, and are explicitly
\begin{align}\label{eqn:cubicsolutions}
p_1 &=\frac{\sqrt[3]{2}
   \left(\sqrt{81 \alpha -12}-9
   \sqrt{\alpha }\right)^{2/3}+2
   \sqrt[3]{3}}{6^{2/3}
   \sqrt[3]{\sqrt{3} \sqrt{\alpha ^3
   (27 \alpha -4)}-9 \alpha
   ^2}},\nonumber\\
p_2 &=\frac{\sqrt[3]{-1}
   \left(\sqrt[3]{-2} \left(\sqrt{81
   \alpha -12}-9 \sqrt{\alpha
   }\right)^{2/3}-2
   \sqrt[3]{3}\right)}{6^{2/3}
   \sqrt{\alpha } \sqrt[3]{\sqrt{81
   \alpha -12}-9 \sqrt{\alpha
   }}},\nonumber\\
p_3 &=\frac{2 (-1)^{2/3}
   \sqrt[3]{3}-\sqrt[3]{-2}
   \left(\sqrt{81 \alpha -12}-9
   \sqrt{\alpha
   }\right)^{2/3}}{6^{2/3}
   \sqrt[3]{\sqrt{3} \sqrt{\alpha ^3
   (27 \alpha -4)}-9 \alpha
   ^2}}.
\end{align}

Depending on the value of $\alpha$, we may have one, two, or no real and positive turning points. In fact, $p_2$ is always a real and negative solution, which must be physically excluded. On the other hand, $p_1$ and $p_3$ are real and positive solutions for $\alpha < 4/27$, representing two turning points. For $\alpha > 4/27$, $p_1$ and $p_3$ become complex conjugate solutions, implying no turning point. For $\alpha = 4/27 = 0.\overline{148}$, these two real solutions merge into a single turning point with $p=3/2$, corresponding to an unstable circular orbit. That is well-known for photons in Schwarzschild space-time, e.g., Eq.\ (11.18) in Ref.~\onlinecite{Schutz2Ed}. 

\begin{figure}[!hb]
	\begin{center}
 	\includegraphics[width=8.5cm]{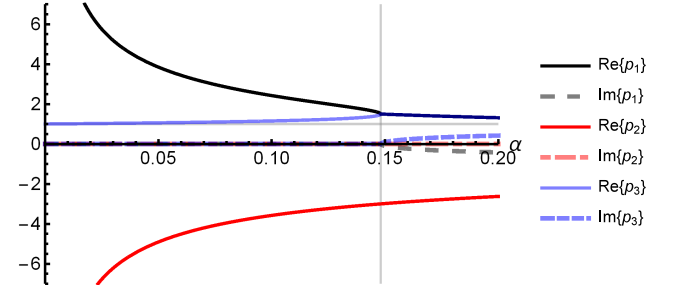}\\
      \caption{\label{fig:cubicsolutions} Solutions of the cubic \Eq{periastra}. The parameter $\alpha$ is defined in \Eq{eqn:alpha}. For $\alpha > 4/27$ there are no positive real solutions. The point $p = 3/2$ at which $p_1$ and $p_3$ coalesce intersects the vertical gray line where $\alpha = 4/27$. Physically, that corresponds to an unstable circular orbit. For decreasing $\alpha$, down to $\alpha \rightarrow \ 0$, $p_3$ monotonically decreases toward $r_S=1$, marked by a horizontal gray line.} 
	\end{center}
\end{figure}

The behavior of the solutions to the cubic \Eq{periastra} is graphed in \Figref{fig:cubicsolutions} as a function of $\alpha$. That behavior is quite consistent with the effective potential for \textit{null} geodesics in Schwarzschild space-time, as shown in Fig.\ 11.2 of Ref.~\onlinecite{Schutz2Ed}, for example. For the `curved-time' metric, \Eq{metricwithoutspatialcurvature}, the corresponding effective potential for \textit{null} geodesics becomes energy-dependent and diverging at $r_S$. However, its basic features do not qualitatively differ from those pertaining to Schwarzschild space-time with regard to the results that we have just provided for turning points of \textit{null} geodesics.

By the same method and procedures that we have applied to study geodesics in Schwarzschild spatial submetric, \Eq{spatialmetric}, or `splittable space-time' metric, \Eq{metricwithouttimecurvature}, equivalently, we have obtained analytic solutions and numerical results for all kinds of geodesics in both the `curved-time' metric, \Eq{metricwithoutspatialcurvature}, and in Schwarzschild's space-time metric, \Eq{metric}. The latter study is critically important, but too extensive to be reported here.\cite{Nick} Therefore, in the remainder of this Section, we will just confine our discussion to comparisons of \textit{null-geodesic asymptotic deflections} for all three metrics considered. 

For the parameters and limit of light grazing the sun, where $r_p = 235,438r_S$, our results indicate a `spatial bending' of half the total GR inward light deflection of 1.75 arc-seconds, which we recover for the exact null geodesic orbit \Eq{nullgeodesicorbitradial} of Schwarzschild space-time. Our results for the null geodesic \Eq{exactnospatialcurvatureradialnullgeodesicorbit} for the `curved-time' metric of \Eq{metricwithoutspatialcurvature} also indicate a `time bending' of half the total GR deflection of 1.75 arc-seconds. Coincidentally, half of the correct GR deflection also agrees with the much older prediction made by Cavendish (1784) and Soldner (1801) based on a purely Newtonian description of light particles: cf.\ Ref.~\onlinecite{Berry}, Sec.\ 5.4, pp.\ 85-88, and Ref.~\onlinecite{Will}. 

\begin{figure}[!hb]
	\begin{center}
 	\includegraphics[width=8.6cm]{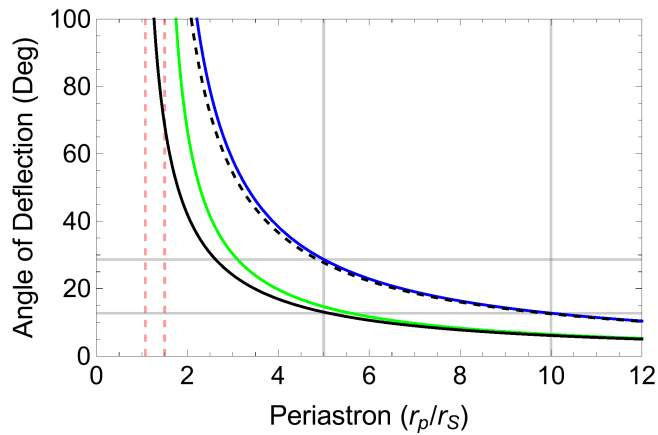}\\
      \caption{\label{fig:angledeflection1} Asymptotic angular deflections vs. periastron for \textit{null} geodesics in all three metrics considered. Black curve is for `spatial bending' with $g_{tt}=-1$. Green curve is for `time bending' with $g_{rr}=1$. Blue curve is for Schwarzschild space-time metric with $g_{tt}*g_{rr}=-1$. The black-dashed curve represents the sum of the black and green curves, i.e., the sum of `spatial bending' and `time bending.' The two red-dashed vertical lines emphasize divergences at the Schwarzschild radius ($r_S = 1$) for `spatial bending' and at the radius of the unstable circular orbit for either `curved-time' or Schwarzschild space-time metrics ($p = 3/2$). Vertical and horizontal gray lines refer to $p=5, 10$ values and to corresponding Schwarzschild space-time deflections, respectively.} 
      \end{center}
\end{figure}

However, for a much closer approach of $r_p$ to $r_S$, `time bending' largely exceeds `spatial bending' of light, while their sum remains substantially below the total GR inward light deflection in Schwarzschild space-time. Some significant values are reported in Table 1. Asymptotic angular deflections vs.\ the periastron for \textit{null} geodesics for all three metrics considered are plotted in \Figref{fig:angledeflection1}.
\begin{table}[h!]
\centering
 \begin{tabular}{|| c | c | c | c | c ||} 
 \hline
 $r_p/r_S$ & Curved Space & Curved Time & Sum & GR Space-Time \\
 \hline\hline
    1.5     & 68.61$^\circ$            & --                        & --                        & -- \\ 
    1.51    & 67.72$^\circ$            & 298.9$^\circ$             & 366.6$^\circ$             & 529.0$^\circ$\\
    1.6     & 60.76$^\circ$            & 152.1$^\circ$             & 212.8$^\circ$             & 274.4$^\circ$\\
    2       & 42.05$^\circ$            & 67.09$^\circ$             & 109.1$^\circ$             & 125.1$^\circ$\\
    5       & 13.04$^\circ$            & 14.65$^\circ$             &  27.69$^\circ$            & 28.66$^\circ$\\
    10      & 6.093$^\circ$            & 6.423$^\circ$             & 12.52$^\circ$             & 12.71$^\circ$\\
    100     & 34.58$^{\prime}$         & 34.75$^{\prime}$          &  1.156$^\circ$            & 1.157$^\circ$\\
    235,438 & 0.876$^{\prime\prime}$   & 0.876$^{\prime\prime}$    &  1.752$^{\prime\prime}$   & 1.752$^{\prime\prime}$ \\
 \hline
 \end{tabular}\caption{Asymptotic angular deflections for some significant periastron values of \textit{null} geodesics in all three metrics considered.}
\end{table}

\section{Conclusions}
We have solved geodesic orbit equations and characterized corresponding manifolds for metrics associated with Schwarzschild geometry, considering space and time curvatures separately. 

For `fixed' or `a-temporal' space, with a positive-definite submetric, \Eq{spatialmetric}, and for an essentially equivalent `splittable space-time' metric, \Eq{metricwithouttimecurvature}, we have provided a central geodesic orbit \Eq{spatialradialgeodesicorbitperiastron}. We have solved that equation in terms of elliptic integrals and functions. The intrinsic geometry of a geodesic equatorial plane with two sides joined at the horizon corresponds to that of a full Flamm's paraboloid. Two kinds of geodesics thus emerge. Both kinds may or may not encircle the hole region any number of times, crossing themselves correspondingly. Regular geodesics reach a periastron $r_p > r_S$, thus remaining confined to a half of Flamm's paraboloid. Singular or $s$-geodesics tangentially reach the $r_S$ circle. These $s$-geodesics must then be regarded as funneling through the $Z=0$ `belt' of the full Flamm's paraboloid. Infinitely many geodesics can possibly be drawn between any two points, but they must be of specific regular or singular types. A precise classification can be made in terms of impact parameters. Geodesic structure and completeness is conveyed by computer-generated figures depicting either Schwarzschild equatorial plane or Flamm's paraboloid.

For the `curved-time' metric of \Eq{metricwithoutspatialcurvature}, devoid of any spatial curvature, geodesic orbits have the same apsides as in Schwarzschild space-time. In particular, apsides of \textit{null} geodesics obey a cubic \Eq{periastra} that we solve. For the parameters and limit of light grazing the sun, asymptotic `spatial bending' and `time bending' become essentially equal, adding up to the total inward light deflection of 1.75 arc-seconds predicted by GR. However, for a much closer approach of $r_p$ to $r_S$, `time bending' largely exceeds `spatial bending' of light, while their sum remains substantially below that of Schwarzschild space-time. These results are exact and generalize or clarify previous statements on that matter.\cite{Price, Ellingson}

\acknowledgments

The authors of this paper are listed in alphabetical order. We acknowledge financial support from NASA/ADAP grants NNH11ZDA001N \& NNX13AI48G and from the Vitreous State Laboratory at the Catholic University of America. We dedicate our work to the memory of Maria Rita Soverchia Resca.

\bibliographystyle{apsrev4-1}


\end{document}